\documentclass[prd,aps,floats,preprint,preprintnumbers,amssymb,nofootinbib]{revtex4}
\usepackage{graphicx}
\usepackage{graphics}
\usepackage{amssymb}
\usepackage{amsthm}
\def\bea{\begin{eqnarray}}
\def\eea{\end{eqnarray}}
\def\bec{\begin{center}}
\def\eec{\end{center}}
\newcommand{\beq}{\begin{equation}}
\newcommand{\eeq}{\end{equation}}
\newcommand{\nn}{\nonumber}

\begin{document}
\title{Vortex Scattering and Intercommuting Cosmic Strings on a Noncommutative Spacetime}
\author{Anosh Joseph}
\affiliation{Department of Physics, Syracuse University, Syracuse, NY 13244 USA}
\author{Mark Trodden}
\affiliation{Center for Particle Cosmology, Department of Physics and Astronomy, University of Pennsylvania, PA 19104, USA}

\begin{abstract}
We study the scattering of noncommutative vortices, based on the noncommutative field theory developed in \cite{Balachandran:2006pi}, as a way to understand the interaction of cosmic strings. In the center-of-mass frame, the effects of noncommutativity vanish, and therefore the reconnection of cosmic strings occurs in an identical manner to the commutative case. However, when scattering occurs in a frame other than the center-of-mass frame, strings still reconnect but the well known 90$^{\circ}$ scattering no longer need correspond to the head on collision of the strings, due to the breakdown of Lorentz invariance in the underlying noncommutative field theory.
\end{abstract}

\date{\today}
\maketitle
\preprint{SU-4252-900}
\section{Introduction}
Topological defects such as magnetic monopoles, cosmic strings and domain walls, arise in a large class of spontaneously broken field theories. More recently, cosmic strings have also been shown to arise within string theory, providing a potential indirect way to search for observational signatures of the theory. The existence of defects often yields tight cosmological constraints, since they have the potential to overclose the universe, to yield nontrivial gravitational wave signatures, or to have nontrivial microphysical interactions. To balance this, there are a number of approaches to standard cosmological problems in which topological defects may play an important role.

Cosmic strings are of particular interest, since their self interactions allow a potentially catastrophic string network to lose energy in an orderly fashion, leading to a scaling solution which need not dominate the universe, and thus may contribute to cosmology in interesting ways. For example, while cosmic strings cannot play the central role in seeding structure formation in the universe, some contribution is still allowed~\cite{Wyman:2005tu} by WMAP and SDSS data, as long as the defects account for no more than 14\% of the temperature fluctuations in the cosmic microwave background radiation. 

Central to an understanding of the cosmological implications of cosmic strings is therefore a detailed understanding of their self interactions. The evolution of cosmic string networks has been thoroughly investigated both numerically and analytically~\cite{Vachaspati:1984dz,Kibble:1984hp, Bennett:1985qt, Bennett:1986zn, Allen:1990tv, Allen:1990mn}. The scattering of cosmic strings exhibits a crucial feature - they reconnect (intercommute or exchange end points) with a probability close to one, after they collide with each other. This property allows large cosmic strings to break down into smaller strings and loops of strings. The loops themselves are (assuming they are non-superconducting) entirely unstable, and shrink to zero size by emitting energy in the form of gravitational radiation and/or Goldstone bosons~\cite{Hawking:1990tx, Fort:1993zb}.

In this paper we investigate the possibility of the reconnection of cosmic strings when the spacetime is noncommutative. It has been suggested that quantum gravity and string theory contain hints that spacetime may be noncommutative at a length scale close to the Planck scale. Given this possibility, it is natural to wonder whether it is possible for noncommutative cosmic strings to reconnect after they collide with each other. 

There exists~\cite{Gopakumar:2000zd, Bak:2000ac, Bak:2000im, Jatkar:2000ei, Bak:2000ym, Lechtenfeld:2001gf, Tong:2002xt, Lindstrom:2000kh} a variety of approaches to constructing and studying the properties of noncommutative solitons. In~\cite{Gopakumar:2000zd} classical stable solitons were constructed for noncommutative scalar field theories, and noncommutative vortex solitons were constructed and studied in~\cite{Bak:2000ac, Bak:2000im, Bak:2000ym, Jatkar:2000ei}. The moduli space dynamics of noncommutative vortices were analyzed in~\cite{Tong:2002xt}, and the scattering of noncommutative solitons was studied in \cite{Lechtenfeld:2001gf, Lindstrom:2000kh}.

In this paper we approach the question of the scattering, and hence reconnection, of cosmic strings by considering the noncommutative abelian Higgs model based on the twisted Poincar\'e symmetry with deformed statistics developed in~\cite{Balachandran:2006pi}. (See~\cite{Chaichian:2004za, Aschieri:2005yw, Balachandran:2005eb, Balachandran:2007kv, Balachandran:2007vx, Akofor:2008ae, Balachandran:2009sq} for more details and developments.) We demonstrate that the nonlocal and Lorentz non-invariant nature of the noncommutative field theory plays a crucial role in the scattering of noncommutative vortices in $2+1$ dimensions, but do not find a significant modification of the behavior of the related cosmic strings in $3+1$ dimensions. The paper is organized as follows. In section~\ref{comm-abelian-higgs} we briefly review the abelian Higgs model in the commutative case. In section~\ref{noncomm-spacetime} we then review the particular formulation of noncommutative field theory that we study, providing a description that we hope will be useful to readers not familiar with this construction. In section~\ref{noncomm-abelian-higgs} we construct the noncommutative abelian Higgs model, and in section~\ref{sec:dynamics} we then discuss the low energy dynamics of noncommutative vortices and describe how their scattering is qualitatively and quantitatively different from that of their commutative counterparts, before concluding. Throughout this paper we use the mostly negative signature.

\section{Vortices in the abelian Higgs Model}
\label{comm-abelian-higgs}
The commutative abelian Higgs model in $d$ spacetime dimensions has Lagrangian density
\beq
\label{eq:lagrangian}
{\cal L} = -\frac{1}{4}F_{\mu \nu} F^{\mu \nu} +  D_{\mu}\phi (D^{\mu}\phi)^{\dagger} - V(\phi) \ ,
\eeq
where $\phi$ is a complex scalar field ($\phi = \phi_1 + i \phi_2$), $A_{\mu}$ is a gauge field charged under the $U(1)$ symmetry and  $\mu, \nu = 0, 1, 2, \cdots, d$. Here, the field strength tensor is defined as $F_{\mu \nu} = \partial_{\mu}A_{\nu} - \partial_{\nu}A_{\mu}$, the covariant derivative $D_{\mu}$ acts as
\beq 
D_{\mu}\phi = (\partial_{\mu} - ig A_{\mu}) \phi
\eeq
and the Higgs potential is
\bea
\label{eq:potential}
V(\phi) &=&\frac{\lambda}{4} (\phi \phi^{\dagger}  - v^2)^2 \ ,
\eea
with $\lambda$ a coupling and $v$ the vacuum expectation value (VEV) of $\phi$ (note that the mass dimensions of the parameters of the theory depend on the total number of dimensions).

Once the local $U(1)$ symmetry is spontaneously broken in vacuum, the field $\phi$ acquires a mass $m_{\phi} = \sqrt{\lambda}v$ and the gauge field $A_{\mu}$ acquires a mass $m_A = \sqrt{2}gv$.

The equations of motion are
\bea
D^{\mu}D_{\mu} \phi &=& \frac{\lambda}{2} (\phi \phi^{\dagger} - v^2)\phi \ ,\\
\partial_{\mu} F^{\mu \nu} &=& -ig [\phi (D^{\nu}\phi)^{\dagger}-(D^{\nu}\phi)\phi^{\dagger}] \ .
\eea
It is convenient to work in the temporal gauge $A_0=0$, in which the equation of motion associated with $A_0$ must be imposed as a constraint (Gauss's law), as
\beq
\partial_i \dot{A}_i + ig [\phi \dot{\phi}^{\dagger} - \dot{\phi}\phi^{\dagger}]= 0 \ .
\eeq

If we now focus on the behavior of vortices in $2+1$ dimensions, and define kinetic and potential energies $T$ and $V$ respectively by
\bea
\label{eq:ke}
T &=& \int d^2x ~\frac{1}{2}\dot{A}_i \dot{A}_i + \dot{\phi}\dot{\phi}^{\dagger} \ ,\\
\label{eq:pe}
V &=& \int d^2x ~D_i\phi (D_i\phi)^{\dagger} + \frac{1}{2} F_{12}^2 + \frac{\lambda}{4}(\phi \phi^{\dagger} - v^2)^2 \ ,
\eea
then the Lagrangian is $L = T - V$, and the total energy $E=T+V$ is a conserved quantity. Its finiteness implies the boundary conditions for the field $\phi$ at spatial infinity.
\beq
\label{eq:boundaryc}
|\phi| \rightarrow v \ , \ \ \ \ D_i\phi \rightarrow 0 \ ,
\eeq
as $|x| \rightarrow \infty$.

When the fields are static, that is, when $\dot{A}_i = 0$, $\dot{\phi} = 0$, the kinetic energy $T$ vanishes, and we may then pose the cylindrically symmetric ansatz 
\bea
\phi(\hat{x}) &=& \rho(r) e^{im \vartheta} \ ,\\
A_i (\hat{x})&=& \alpha(r)\hat{\vartheta}
\eea
characterizing a vortex of winding number $m$. Our criteria of finite energy per unit length and regularity at the origin then yield the boundary conditions $\rho(r) \rightarrow v$ and $\alpha(r) \rightarrow 1/gv$ as $r \rightarrow \infty$; and $\rho(r) \rightarrow 0$, $\alpha(r) \rightarrow 0$ as $r \rightarrow 0$. The corresponding solution is the commutative abelian Higgs vortex, and if we add in an extra spatial dimension, along which the configuration is translationally invariant, then the solution describes the $3+1$ dimensional cosmic string. 

For simplicity, in this paper, we focus on vortices at the Bogomol'nyi self-dual point, for which the coupling takes the critical value $\lambda = 2g^2$. In this case, the masses are equal, $m_{\phi} = m_{A}$, the forces between the vortices vanish and it is possible to find stable static multivortex configurations. 

\section{Noncommutative Spacetime and Deformed Poincar\'e Symmetry}
\label{noncomm-spacetime}

In the next section we will construct the noncommutative analogue to the abelian Higgs model. In order to do so, we will need to lay out  precisely what we mean by a noncommutative spacetime. We will work on the {\it Moyal spacetime} defined by the algebra \cite{Connes, Madore, Landi}
\beq
\label{eq:comm}
[\hat{x}_{\mu}, \hat{x}_{\nu}] = i \theta_{\mu \nu}{\mathbb I} \ ,
\eeq
where the coordinate operators $\widehat{x}_{\mu}$ yield the Cartesian coordinates $x_{\mu}$ of (flat) spacetime via $\widehat{x}_{\mu} (x) = x_{\mu}$, and $\theta_{\mu \nu} = - \theta_{\nu \mu}$ are constants. In the limit $\theta_{\mu \nu} \rightarrow 0$, one recovers ordinary commutative spacetime.

Operator valued functions on the Moyal spacetime form a noncommutative algebra ${\cal A}_{\theta}$, the elements of which can be identified with ordinary functions on ${\mathbb R}^4$, with the product of two functions, $f$ and $g$ say, given by the Moyal product ($\star$-product)
\beq
f \star g (x) = \exp\Big[\frac{i}{2} \theta^{ij} \frac{\partial}{\partial x_1^i} \frac{\partial}{\partial x_2^i}\Big] f(x_1) g(x_2)\Big|_{x_1 = x_2 = x} \ .
\eeq

The commutation relations~(\ref{eq:comm}) are not invariant under the usual Lorentz transformations, and so the Lorentz symmetry is broken. However, it is possible to impose invariance under a deformed Lorentz Symmetry~\cite{Chaichian:2004za,Balachandran:2006pi, Aschieri:2005yw, Balachandran:2005eb} as we briefly explain in appendix~\ref{appendixA}.

The noncommutative field $\varphi_{\theta}$ differs form its commutative counterpart $\varphi$ in two ways: $i$.) It belongs to the noncommutative algebra of functions on Minkowski spacetime ${\cal M}^{4}$ and $ii$.) it obeys deformed statistics. The deformed statistics can be accounted for by writing
\beq
\label{eq:twistedfield}
\varphi_{\theta} = \varphi~\textrm{e}^{\frac{1}{2}\overleftarrow{\partial} \wedge P}
\eeq
where $\overleftarrow{\partial} \wedge P \equiv \overleftarrow{\partial}_{\mu}\theta^{\mu \nu}P_{\nu}$ and $P_{\mu}$ is the total momentum operator for all the fields.

From this it follows that the $\star$-product of an arbitrary number of fields $\varphi_{\theta}^{(i)}$ ($i$ = 1, 2, 3, $\cdots$) is
\beq
\varphi_{\theta}^{(1)} \star \varphi_{\theta}^{(2)} \star {\cdots} =(\varphi^{(1)}\varphi^{(2)} {\cdots})~e^{\frac{1}{2} \overleftarrow{\partial} \wedge P}.
\label{eq:productfields}
\eeq
Although the rule (\ref{eq:twistedfield}) is for a massive scalar field, it also applies to all bosonic and Grassmann-valued matter fields.

Matter fields on ${\cal A}_{\theta}(\mathbb{R}^{4})$ must be transported by the connection compatibly with~(\ref{eq:twistedfield}), and therefore a natural choice for the covariant derivative is~\cite{Balachandran:2007kv}
\beq
D_{\mu} \varphi_{\theta} = (D_{\mu}^{c} \varphi)~e^{\frac{1}{2}\overleftarrow{\partial} \wedge P},
\label{eq:covariant}
\eeq
where
\beq
D_{\mu}^{c} \varphi = \partial_{\mu}\varphi - igA_{\mu}\varphi \ ,
\eeq
and we define $A_{\mu}\varphi(x)\equiv A_{\mu}(x)\varphi(x)$ to mean point-wise multiplication. This can also be written using the $\star$-product as
\beq
D_{\mu}\varphi_{\theta} = \Big(D_{\mu}^{c} e^{\frac{1}{2} \overleftarrow{\partial} \wedge
P}\Big)\star \Big(\varphi e^{\frac{1}{2} \overleftarrow{\partial} \wedge
P}\Big) \ .
\eeq

This choice of $D_{\mu}$ preserves statistics, Poincar\'e and gauge invariance, and the requirement that $D_{\mu}$ is associated with the commutative algebra ${\cal A}(\mathbb{R}^{N})$~\cite{Balachandran:2007kv}
\begin{eqnarray}
[D_{\mu}, D_{\nu}] \varphi_{\theta} &=& \Big([D^{c}_{\mu}, D^{c}_{\nu}]\varphi\Big)e^{\frac{1}{2}\overleftarrow{\partial} \wedge P} \\
&=&\Big(F_{\mu \nu}^{c}\varphi\Big)e^{\frac{1}{2}\overleftarrow{\partial} \wedge
P}.
\end{eqnarray}
As $F_{\mu \nu}^{c}$ is the standard $\theta^{\mu \nu}=0$ field strength tensor, our gauge field is associated with ${\cal A}(\mathbb{R}^{N})$. This lays out the components of the Moyal spacetime necessary for our analysis. A complete description of the gauge theory formulation we adopt here can be found in~\cite{Balachandran:2007kv, Balachandran:2007vx, Akofor:2008ae}.

\section{The Noncommutative Abelian Higgs Model}
\label{noncomm-abelian-higgs}

The noncommutative abelian Higgs model is constructed by replacing the ordinary pointwise multiplication between the fields by a Moyal product and identifying the noncommutative fields as statistics-deformed fields. The Lagrangian density is
\beq
\label{eq:L}
{\cal L} = -\frac{1}{4}F_{\mu \nu} \star F^{\mu \nu} +  D_{\mu}\phi_{\theta} \star (D^{\mu}\phi_{\theta})^{\dagger}  - V_{\star}(\phi_{\theta}) \ ,
\eeq
with $F_{\mu \nu} \equiv F^c_{\mu \nu}$ and $D_{\mu} \equiv D^c_{\mu}= \partial_{\mu} - ig A_{\mu}$.

The Higgs potential term takes the following form in terms of the associated commutative field
\bea
V_{\star}(\phi_{\theta}) &=&\frac{\lambda}{4} (\phi_{\theta} \star \phi_{\theta}^{\dagger}  - v^2)_{\star}^2 \nn \\
&=& \frac{\lambda}{4}(\phi^{\dagger} \phi - v^2) e^{\frac{1}{2} \overleftarrow{\partial} \wedge P}.
\eea

As in the commutative case, it is convenient to work in the temporal gauge $A_0=0$, in which the Gauss' law constraint becomes
\beq
\Big(\partial_i\dot{A}_i+ ig [\phi \dot{\phi}^{\dagger} - \dot{\phi}\phi^{\dagger}]\Big)e^{\frac{1}{2} \overleftarrow{\partial} \wedge P}= 0 \ .
\eeq
The Lagrangian can then once again be written in the form $L = T - V$, where $T$ and $V$ are the kinetic and potential energies, given by
\bea
\label{eq:kenc}
T &=& \int d^2x ~\frac{1}{2}\dot{A}_i \star \dot{A}_i + \dot{\phi}_{\theta} \star \dot{\phi}^{\dagger}_{\theta} \ ,\\
\label{eq:penc}
V &=& \int d^2x ~(D_i\phi_{\theta})^{\dagger} \star D_i\phi_{\theta}  + \frac{1}{2} F_{12} \star F_{12} + \frac{\lambda}{4}(\phi_{\theta}^{\dagger} \star \phi_{\theta}  - v^2)_{\star}^2 \ .
\eea
Here we have used $\star$-multiplication even between the terms involving the gauge fields, since the spontaneous breakdown of the $U(1)$ symmetry makes the gauge field a massive gauge boson. 

Without loss of generality we choose the third spatial direction to commute with the other two spatial directions. Then, representing the Moyal product in terms of the commutative fields and the exponential involving the momentum operator, we note that the spatial integration removes the spatial part of the derivative in the exponential, which appears as a surface term.  Thus the kinetic and potential energies take the form
\bea
\label{eq:kenc1}
T &=& \int d^2x ~\Big(\frac{1}{2}\dot{A}_i \dot{A}_i + \dot{\phi}\dot{\phi}^{\dagger}\Big)e^{\frac{1}{2}\overleftarrow{\partial}_0 \theta^{0i} P_i} \ ,\\
\label{eq:penc1}
V &=& \int d^2x ~\Big((D_i\phi)^{\dagger} D_i\phi  + \frac{1}{2} F_{12} F_{12} + \frac{\lambda}{4}(\phi^{\dagger} \phi  - v^2)^2\Big) e^{\frac{1}{2}\overleftarrow{\partial}_0 \theta^{0i} P_i} \ .
\eea

One result is then immediately clear. In the static case, the effect of the noncommutativity entirely vanishes, since $P_i=0$. Thus, in the static case, the analysis follows the commutative case, and the structure of noncommutative vortices is the same as their commutative counterparts. However, as we shall see, in the case of moving vortices it is necessary to include the effect of noncommutativity, and the factor $ e^{\frac{1}{2}\overleftarrow{\partial}_0 \theta^{0i} P_i}$ becomes relevant.

\section{Low energy Dynamics: The Geodesic Approximation}
\label{sec:dynamics}
\subsection{Commutative Case}

The abelian Higgs model at the Bogomol'nyi self-dual point saturates a topological lower bound on the field energy and admits static multivortex configurations. The low energy dynamics of multivortex solutions may then be approximated by motion on the space of corresponding static solutions~\cite{Manton:1981mp}. 

If ${\cal C}$ is the space of field configurations of the theory, then the $n$-vortex solutions form a submanifold $M_n$, called the moduli space, of ${\cal C}$ on which the potential energy $V$ takes its absolute minimum. Imparting a small kinetic energy to the field configuration corresponds to a slow motion tangent to $M_n$. In the subsequent evolution of the field configuration, the trajectory of the system will be constrained by $V$ to lie close to $M_n$. Thus, $V$ remains approximately constant, and the field evolution is described by geodesic motion on $M_n$, the metric being induced by the kinetic energy Lagrangian $T$. The problem of describing the vortex dynamics is thus reduced to finding the metric and solving the ordinary differential geodesic equations on $M_n$. For a detailed description of the low energy vortex dynamics and scattering in the geodesic approximation for the commutative case, we refer the reader to~\cite{Samols:1991ne, Ruback:1988ba, Myers:1991yh}.

We now focus on two slowly moving identical vortices, for which the moduli space is $M_2$. Since the vortex dynamics is happening on the plane ${\mathbb R}^2$, it is useful to make the identification ${\mathbb R}^2 \simeq {\mathbb C}$ and write the position of a point ($x_1$, $x_2$) on ${\mathbb R}^2$ as $z=x_1 + ix_2$. We also use the complex notation $A = \frac{1}{2}(A_1 +i A_2)$. The kinetic energy Lagrangian, in terms of $A$ and $\phi$, is
\beq
T = \int d^2 x~(2 \dot{A}\dot{\bar{A}} + \dot{\phi}\dot{\bar{\phi}}) \ .
\eeq

For the case of two vortices this can be reduced to the following form~\cite{Samols:1991ne}
\beq
\label{eq:T-reduced}
T = \pi v^2 \sum_{r,s =1}^{2} \Big(\delta_{rs} + 2 \frac{\partial \bar{h}_s}{\partial z_r}\Big)\dot{z}_r \dot{\bar{z}}_s \ ,
\eeq
in which $\pi v^2$ is the static energy of a single vortex, $z_k$ represent the locations of vortices (zeros of the Higgs field) on the plane, and $h_s$ is a complex valued function.

The above expression for the kinetic energy leads to the metric 
\beq
\label{eq:metric}
ds^2 = \sum_{r,s =1}^{2} \Big(\delta_{rs} + 2 \frac{\partial \bar{h}_s}{\partial z_r}\Big)dz_r d{\bar{z}}_s
\eeq
appropriate for use in the geodesic approximation. Here we have chosen to normalize the metric relative to $T$ by dividing by the single vortex energy $\pi v^2$.

Since the parent field theory (\ref{eq:lagrangian}) is invariant under translations and rotations on the plane ${\mathbb R}^2$, the vortex metric also inherits that property. And since translational invariance implies the conservation of linear momentum $P = P_1 +i P_2 = \pi v^2 \sum_{r=1}^{2} \dot{z}_r$, an immediate consequence is that we may analyze the two-vortex system in the center-of-mass coordinates.

On using the center-of-mass and relative coordinates $Z = \frac{1}{2}(z_1 + z_2)$, $\xi_1 = -\xi_2 = \xi \equiv \frac{1}{2} (z_1-z_2)$ respectively, the metric (\ref{eq:metric}) takes the form
\beq
\label{eq:metric2}
ds^2 = 2 dZ d\bar{Z} + \sum_{r,s =1}^{2} \Big(\delta_{rs} + 2 \frac{\partial \bar{h}_s}{\partial z_r}\Big)d\xi_r d{\bar{\xi}}_s \ .
\eeq
Since the parent theory is symmetric under $\phi \rightarrow -\phi$, this implies the constraint $h_1 = -h_2$. Thus the expression for the metric (\ref{eq:metric2}) then reduces to
\beq
ds^2 = 2 dZ d\bar{Z} + \Big(1 + 2 \frac{\partial \bar{h}_1}{\partial \xi}\Big)d\xi d{\bar{\xi}} \ .
\eeq

We introduce polar coordinates ($\rho$, $\vartheta$) defined by 
\beq
\xi = \rho e^{i \vartheta}
\eeq
where the ranges of $\rho$ and $\vartheta$ are: $0 \leq \rho < \infty$ and $-\frac{\pi}{2} \leq \vartheta \leq \frac{\pi}{2}$. For a fixed $Z$, $\xi$ and $-\xi$ label the same point in moduli space and should be identified. That is, we should identify $\vartheta = -\pi/2$ and $\vartheta = \pi/2$. 

Since the center-of-mass system is symmetric under rotations and reflections, we may write $h_1 = h(\rho) e^{-i \vartheta}$, with $h(\rho)$ real, so that the metric describing the relative motion is~\cite{Samols:1991ne}
\beq
ds_{\rm{rel}}^2 = \frac{1}{2}F^2(\rho) (d\rho^2 + \rho^2 d\vartheta^2) \ .
\eeq
This reduction to just a single unknown function $F(\rho)$ is a consequence of the hermiticity of the metric, which itself is inherited from the reality of the kinetic energy $T$ which, in units of the static vortex energy $\pi v^2$, reduces to
\beq
\label{eq:kineticl}
T(\rho, \vartheta) = \frac{1}{2} F(\rho)(\dot{\rho}^2 + \rho^2 \dot{\vartheta}^2) \ .
\eeq

\begin{figure}
\includegraphics[height=8cm]{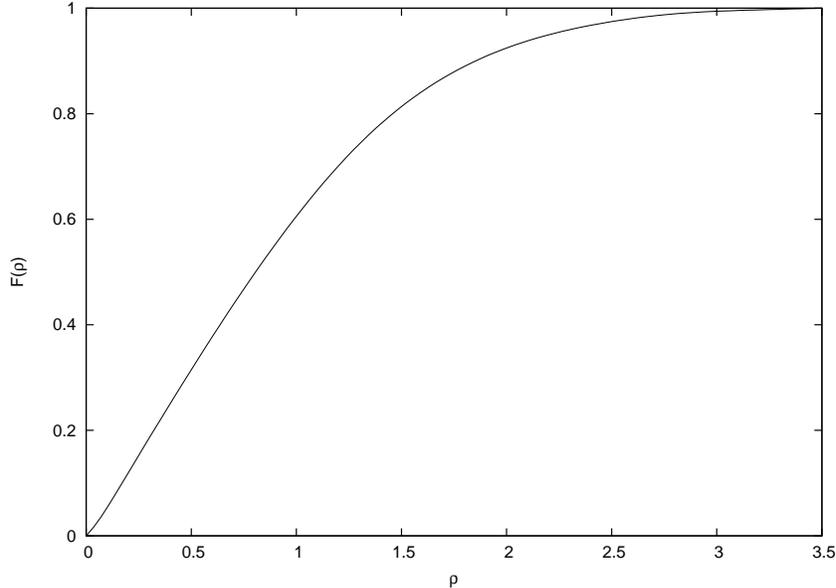}               
\caption{The profile of $F(\rho)$ in the commutative case~\cite{Samols:1991ne}.}
\label{fig:F-rho}
\end{figure}

The function $F(\rho)$ depends only on the relative separation of the vortices, and should go to zero as the two vortices begin to overlap. Samols has calculated $F(\rho)$ numerically~\cite{Samols:1991ne} and we display his results in figure~(\ref{fig:F-rho}).

Using the two conserved quantities of the system - the energy $E$ and the angular momentum $l$ - one may derive an equation for $d\rho/d \vartheta$ and integrate to obtain the scattering angle as a function of the impact parameter $b$. This yields~\cite{Myers:1991yh}
\beq
\label{eq:scattering-angle}
\vartheta_{\rm{sc}} (b) = \int_{\rho_0}^{\infty} \frac{2b~d\rho}{\rho\sqrt{\rho^2F^2(\rho)-b^2}} \ ,
\eeq
where $\rho_0$ is the the turning point, given by the solution to $\rho_0 F(\rho_0) = b$.

\subsection{The Noncommutative Case}

We now extend this analysis to the noncommutative case. For two identical vortices the kinetic term~(\ref{eq:kenc1}) can be written as
\beq
\label{eq:kenc2}
T^{(\theta)} = \int d^2x ~\frac{1}{2}\dot{A}_i \dot{A}_ie^{\frac{1}{2}\overleftarrow{\partial}^{(A)}_0 \theta^{0i} P_i} + \dot{\phi}\dot{\phi}^{\dagger} e^{\frac{1}{2}\overleftarrow{\partial}^{(\phi)}_0 \theta^{0i} P_i} \ .
\eeq
In the commutative case the expression for $T$ is manifestly real~\cite{Samols:1991ne}, and so we consider only the real part of~(\ref{eq:kenc2}), yielding
\beq
\label{eq:realKL}
T^{(\theta)} = \int d^2x ~\frac{1}{2} \dot{A}_i \dot{A}_i \cos\Big(\frac{1}{2} \overleftarrow{P}^{(A)}_0 \theta^{0i} P_i\Big) + \dot{\phi}\dot{\phi}^{\dagger}  \cos\Big(\frac{1}{2} \overleftarrow{P}^{(\phi)}_0 \theta^{0i} P_i\Big) \ .
\eeq

As we are dealing with two identical vortices, the initial configuration is given by the ansatz
\bea
A_i(x) &=& A_i^1(x) + A_i^2(x) \ , \ \ \ \ \ \phi(x) = \phi^1(x)\phi^2(x) \ ,
\eea
where the superscripts refer to the two vortices. This ansatz is an excellent approximation when the vortices are separated by distances well in excess of their finite size cores~\cite{Myers:1991yh}.

It is clear from the expression~(\ref{eq:realKL}) that the effect of noncommutativity depends on the combination $\vec{\theta^0} \cdot \vec{P}$, where $\vec{\theta^0} = (\theta^{01}, \theta^{02}, \theta^{03})$ and $\vec{P} = \vec{P}_{inc}$ is the total incident momentum of the scattering vortices. In particular, the phase factors contain $m_A\vec{\theta^0} \cdot \vec{P}_{inc}$ and $m_{\phi}\vec{\theta^0} \cdot \vec{P}_{inc}$ for the (massive) gauge boson $A_{\mu}$ and scalar field $\phi$ respectively. 

At the Bogomol'nyi self-dual point $\lambda = 2g^2$, at which $m_A = m_{\phi}=\sqrt{2}gv$, the kinetic Lagrangian takes the form
\beq
\label{eq:ncT}
T^{(\theta)} = \int d^2x ~\Big(\frac{1}{2} \dot{A}_i \dot{A}_i + \dot{\phi}^{\dagger} \dot{\phi}\Big) \cos \Big(\frac{1}{2}(\sqrt{2}gv) \vec{\theta^0} \cdot \vec{P}_{inc}\Big) \ .
\eeq

Working again in the polar coordinates ($\rho, \vartheta$), the simple noncommutative extension of the kinetic Lagrangian~(\ref{eq:kineticl}) is then
\beq
T^{(\theta)}(\rho, \vartheta) = \frac{1}{2} F^{(\theta)}(\rho)(\dot{\rho}^2 + \rho^2 \dot{\vartheta}^2) \ ,
\eeq
where
\beq
F^{(\theta)}(\rho) = F(\rho)\cos \Big(\frac{1}{2}(\sqrt{2}gv) \vec{\theta^0} \cdot \vec{P}_{inc}\Big),~~F(\rho) \equiv F^{(\theta=0)}(\rho) \ .
\eeq

Notice that this expression has a smooth commutative limit, and the effect of noncommutativity vanishes for the cases $i.)$ when the vectors $\vec{\theta^0}$ and  $\vec{P}_{inc}$ are perpendicular to each other or $ii.)$ when $\vec{P}_{inc}$ vanishes (i.e. when the vortices are in the center-of-mass frame) or $iii.)$ when $\frac{1}{2}(\sqrt{2}gv) \vec{\theta^0} \cdot \vec{P}_{inc} = 2n\pi,~n \in {\mathbb Z}$. It should be noted that in this third case one obtains $F^{(\theta)}(\rho) \rightarrow \pm F(\rho)$ due to the oscillatory nature of the cosine function. Since we are focusing only on the low energy dynamics, where the total momentum is close to zero and the geodesic approximation is valid, we ignore the case in which the sign of $F(\rho)$ is negative. 

However, it is important to realize that this scattering analysis is done in the center-of-mass frame. This implies that $\vec{P}_{inc} = 0$ and consequently there is no effect due to noncommutativity in the scattering process. In the commutative case, it has been shown that vortices scatter at $90^0$ angle at zero impact parameter (head-on collision). The corresponding three-dimensional picture is that of two colliding cosmic strings. Also in the commutative case, two colliding cosmic strings reconnect (exchange end points) after collision. Reconnection of the colliding cosmic strings can be understood as a collection of colliding vortices in two-dimensions with various impact parameters. Thus at the spatial slice with impact parameter $b=0$, the vortex string reconnection is equivalent to the right-angle scattering of the vortices. 

\begin{figure}
\includegraphics[width=.4\textwidth, angle=270]{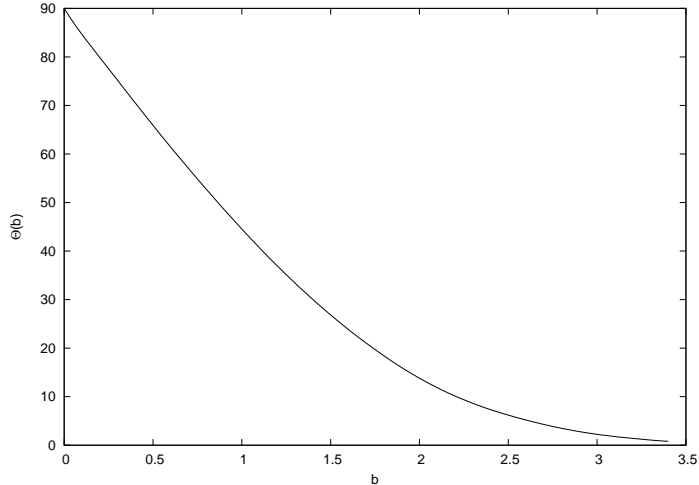}
\caption{The scattering angle $\vartheta_{\rm{sc}}$ as a function of impact parameter $b$ \cite{Samols:1991ne}.}
\label{fig:scatter}
\end{figure}

The simple conclusion we can draw here, consistent with our earlier results, is that two colliding cosmic strings reconnect after collision in the center-of-mass frame even in the noncommutative Moyal spacetime. In figure~(\ref{fig:scatter}) the scattering angle $\Theta$ is plotted as a function of impact parameter $b$ for the commutative case. The vortices scatter at right angles at zero impact parameter in this case. 

Moving away from the center-of-mass frame, we now see that the effect of noncommutativity appears in the scattering analysis through the term $\vec{\theta^0} \cdot \vec{P}_{inc}$. From (\ref{eq:T-reduced}) and (\ref{eq:ncT}), in a non-center-of-mass frame the noncommutative kinetic Lagrangian takes the form
\beq
\label{eq:ncT2}
T^{(\theta)} = \Big\{\pi v^2 \sum_{r,s =1}^{2} \Big(\delta_{rs} + 2 \frac{\partial \bar{h}_s}{\partial z_r}\Big)\dot{z}_r \dot{\bar{z}}_s\Big\} ~\cos \Big(\frac{1}{2}(\sqrt{2}gv) \vec{\theta^0} \cdot \vec{P}_{inc}\Big) \ .
\eeq
In this case it is not possible to reduce (\ref{eq:ncT2}) to a form involving a single function of the relative coordinates as we did in (\ref{eq:kineticl}), since the rotation and reflection symmetries are absent in a non-center-of mass system.

Nevertheless, we can still conclude that two vortices intercommute in a non-center-of mass system, as the intercommutation property of vortices is frame independent. What is different here is that the scattering angle of $90^{\circ}$ (this corresponds to a $180^{\circ}$ scattering in a lab frame, which is a non-center-of-mass frame) may not correspond to the case of zero impact parameter, due to the presence of noncommutativity. Thus the scattering properties of noncommutative vortices are different from those of commutative vortices. This striking feature of noncommutative votex scattering is due to the inherent Lorentz noninvariance of noncommutative field theories. 

\section{Conclusions}

In this paper we have investigated the scattering of noncommutative vortices, and hence the interaction between noncommutative cosmic strings, with the goal of understanding how these may differ from their commutative counterparts. We have worked in the Moyal spacetime, have implemented the effects of noncommutativity by using the star product and by rewriting the ordinary fields as statistics deformed fields, and have focused on the noncommutative version of the abelian Higgs model. We have also used the geodesic approximation to probe the low energy dynamics of vortices, which allows us to express the relevant quantities in terms of the kinetic Lagrangian. 

We have demonstrated several results, the first of which is that noncommutative cosmic strings reconnect after collision, just like their commutative relatives. The effects of noncommutativity in the Moyal spacetime can be captured through operators involving the total momentum operator. This allows us to show, within the geodesic approximation, in which we can phrase the relevant questions in terms of the kinetic Lagrangian, that in the center-of-mass frame the scattering of noncommutative cosmic strings is the same as in of the commutative case. 

In non-center-of-mass frames, however, our formalism allows us to easily see that the scattering of noncommutative vortices can be somewhat different than in the commutative limit. While it is clear that cosmic strings will still reconnect after collision, unlike in the commutative case the well known 90$^{\circ}$ scattering may not correspond to a zero impact parameter collision. Thus, the scattering of noncommutative vortices in $2+1$ dimensions can be seen to be quantitatively different from the commutative case, but the overall behavior of cosmic strings in $3+1$ dimensions remains essentially unchanged by the addition of noncommutativity.

\section{acknowledgments}

We thank Dongsu Bak for useful discussions on noncommutative vortex solitons. AJ's work is supported in part by the US Department of Energy grant under the contract number DE-FG02-85ER40231. The work of MT was supported in part by National Science Foundation grant PHY-0930521, by Department of Energy grant DE-FG05-95ER40893-A020 and by NASA ATP grant NNX08AH27G.

\appendix

\section{The Moyal Spacetime with Twisted Poincar\'e Symmetry and Deformed Statistics.}
\label{appendixA}
Here we briefly discuss the implementation of the twisted Poincar\'e group action compatible with the noncommutative spacetime relations given in (\ref{eq:comm}) and how this gives rise to deformed statistics of the fields. 

\subsection{Twisted Poincar\'e Symmetry}

The Lie algebra ${\cal P}$ of the Poincar\'e group has generators (basis) $M_{\alpha \beta}$ and $P_{\mu}$. The abelian subalgebra of infinitesimal generators $P_{\mu}$ can be used to construct a twist element \cite{drinfeld, chaichian2, chari}
\beq
{\cal F}_{\theta} = \textrm{exp}(-\frac{i}{2}\theta^{\alpha \beta}P_{\alpha} \otimes P_{\beta}), ~~~P_{\alpha} = -i \partial_{\alpha} \ .
\eeq 
(The Minkowski metric with signature ($+, -, -, -$) is used to raise and lower the indices.) This twist element can be used to deform the coproduct, a symmetric map from the universal enveloping algebra ${\cal U}({\cal P})$ of the Poincar\'e algebra to ${\cal U}({\cal P}) \otimes {\cal U}({\cal P})$, in such a way that it is compatible with the above commutation relations.

The coproduct $\Delta_{0}$ appropriate for $\theta_{\mu \nu} =0$ defines the action of ${\cal P}$ on the tensor product of representations. In the case of the generators $X$ of ${\cal P}$, this standard coproduct is
\beq
\Delta_{0}(X) = 1 \otimes X + X \otimes 1 \ .
\eeq

In the presence of the twist, the coproduct $\Delta_{0}$ is modified to $\Delta_{\theta}$ where
\beq
\Delta_{\theta} = {\cal F}_{\theta}^{-1} \Delta_{0} {\cal F}_{\theta} \ .
\eeq

The algebra ${\cal A}_{0}$ of functions on Minkowski space ${\cal M}^{4}$ is commutative with the commutative multiplication $m_{0}$:
\beq
m_{0} (f \otimes g)(x) = f(x)g(x) \ .
\eeq

The Poincar\'e algebra acts on ${\cal A}_{0}$ in a well-known way
\bea
P_{\mu}f(x) &=& -i \partial_{\mu}f(x) \ ,\\
M_{\mu \nu}~f(x) &=& -i(x_{\mu}\partial_{\nu} - x_{\nu}\partial_{\mu})f(x) \ ,
\eea
and acts on tensor products $f \otimes g$ using the coproduct $\Delta_{0}(X)$.

In the Moyal algebra ${\cal A}_{\theta}$, commutative multiplication is changed from $m_0$ to $m_{\theta}$, in terms of which the Moyal $\star$-product can be recast as
\beq
f \star g (x) = m_{\theta} (f\otimes g)(x) = m_{0} ({\cal F}_{\theta}~(f \otimes g))(x) \ .
\eeq

This $\star$-multiplication precisely implements noncommutativity, since it can be shown that it implies~(\ref{eq:comm}):
\beq
\label{eq:comm2}
[\hat{x}_{\mu}, \hat{x}_{\nu}]_{\star} = m_{\theta}(\hat{x}_{\mu}\hat{x}_{\nu} - \hat{x}_{\nu}\hat{x}_{\mu})= i \theta_{\mu \nu}{\mathbb I} \ .
\eeq

Thus, the Poincar\'e algebra acts on functions $f \in {\cal A}_{\theta}$ in the usual way while it acts on tensor products $f \otimes g \in {\cal A}_{\theta} \otimes {\cal A}_{\theta}$ using the coproduct $\Delta_{\theta}(X)$ \cite{Chaichian:2004za, Aschieri}.

\subsection{Deformed Statistics}

It can be shown immediately that the action of the deformed coproduct is not compatible with standard statistics \cite{Balachandran:2006pi, Balachandran:2005eb}. In the commutative case, $\theta^{\mu \nu} =0$, for two scalar fields $\phi^{'}$ and $\phi^{''}$ the exchange operation 
\beq
\varphi^{'} \otimes \varphi^{''} \longrightarrow \varphi^{''} \otimes \varphi^{'} 
\eeq
must not be affected by the Lorentz group action. If we denote the exchange operation by $\tau_0$, we have
\beq
\tau_{0} \Delta_{0}(\Lambda) = \Delta_{0}(\Lambda)\tau_{0},
\eeq
where $\Lambda \in {\cal P}^{\uparrow}_{+}$, the connected component of the Poincar\'e group.

Now since $\tau_{0} {\cal F}_{\theta} = {\cal F}^{-1}_{\theta} \tau_{0}$, we have
\beq
\tau_{0} \Delta_{\theta}(\Lambda) \neq \Delta_{\theta}(\Lambda) \tau_{0} \ ,
\eeq
showing that the use of the usual exchange operation (statistics) is not compatible with the deformed coproduct.

However, if we replace $\tau_0$ by a deformed version, $\tau_{\theta}$, given by
\beq
\tau_{\theta} \equiv {\cal F}^{-1}_{\theta} \tau_{0} {\cal F}_{\theta},~~\tau_{\theta}^{2} = 1 \otimes 1 \ ,
\eeq
then the exchange operation is compatible with the deformed coproduct of the Poincar\'e group.

Thus noncommutative fields have deformed statistics. They obey deformed symmetrization (anti-symmetrization), defined by
\bea
\phi' \otimes_{S_{\theta}, A_{\theta}} \phi'' &\equiv& \Big(\frac{1 \pm \tau_{\theta}}{2}\Big) (\phi' \otimes \phi'') \ ,
\eea
where the `+' sign is for bosonic fields and `-' sign is for Grassman-valued spinor fields.

\end{document}